# Security of mobile agents: a new concept of the integrity protection

Aneta Zwierko[1], Zbigniew Kotulski[1,2]
[1]Instytut Telekomunikacji, WEiTI, Politechnika Warszawska
[2]Instytut Podstawowych Problemów Techniki, PAN
*azwierko@tele.pw.edu.pl*, *zkotulsk@ippt.gov.pl*, zkotulsk@tele.pw.edu.pl

**Abstract**
The recent developments in the mobile technology (mobile phones, middleware) created a need for new methods of protecting the code transmitted through the network. The proposed mechanisms not only secure the compiled program, but also the data, that can be gathered during its "journey". The oldest and the simplest methods are more concentrated on integrity of the code itself and on the detection of unauthorized manipulation. Other, more advanced proposals protect not only the code but also the execution state and the collected data. The paper is divided into two parts. The first one is mostly devoted to different methods of securing the code and protecting its integrity; starting from watermarking and fingerprinting, up to methods designed specially for mobile agent systems: encrypted function, cryptographic traces, time limited black-box security, chained-MAC protocol, publicly-verifiable chained digital signatures The second part presents new concept for providing mobile agents with integrity protection, based on a zero-knowledge proof system.

## 1. Introduction

The mobile agent systems offer new possibilities for the e-commerce applications: creating new types of electronic ventures from e-shops, e-auctions to virtual enterprises and e-marketplaces. Utilizing the agent system helps to automate many electronic commerce tasks. Beyond simple information gathering tasks, mobile agents can take over all tasks of commercial transactions, namely price negotiation, contract signing and delivery of (electronic) goods and services. Such systems are developed for diverse business areas, e.g., contract negotiations, service brokering, stock trading and many others ([4], [11], and [10]).

Mobile agent systems have many advantages over traditional (static) distributed computing environments:
- require less network bandwidth,
- increase asynchrony among clients and servers,
- dynamically update server interfaces,
- introduce concurency.

The benefits from utilizing the mobile agents in various business areas are great. However, this technology brings some serious security risks: one of the most important is the possibility of a tampering an agent. In the mobile agent systems the agent's code and internal data autonomously migrate between hosts and could be easy changed during the transmission or at a malicious host site. An agent cannot itself prevent this, but different countermeasures can be utilized in order to detect any manipulation made by an unauthorized party. They can be integrated directly into the agent system, or only into the design of an agent to extend the capabilities of the underlying agent system.

Several degrees of agent's mobility exist, corresponding to the existing possibilities of relocating code and state information, including the values of instance variables, the program counter, execution stack, etc. The mobile technologies can be divided in to two groups:
- weakly mobile: only the code is migrating no execution state is sent along with an agent program
- strong mobile: a running program is moving to another execution environment (along with its particular state).

In this paper we discuss the agent system mobile in the strong sense.

**Organization of this paper.** First we present basic definitions and notions, which are later used in the description of different methods. Then, we briefly survey the known techniques for protecting agent's integrity. In the Section 5 we present new concept for preventing agent's tampering based on a zero-knowledge proof system. Finally, we present the conclusions and future research areas.

## 2. Definitions and notions

This section presents basic notions concerning agent's integrity that will be later used in description of various solutions (most of the definitions come from [6]).

The integrity of an agent means that neither its code nor execution state can be changed by an unauthorized party or such changes should be detectable (by an owner, a host or an agent platform which wants to interact with the agent).

The authorized changes occur only when the agent have to migrate from one host to another. Below is a more formal definition:

*Definition 1 (integrity of an agent):* An agent's integrity is not compromised if any unauthorized modification can be detected by the agent's owner.

The concept of *forward integrity* is also used in evaluation of many methods. This notion is used in a system where agent's data can be represented as a chain of partial results (a sequence of static pieces of data). *Forward integrity* can be divided into the two types, which differ in their possibility to resist cooperating malicious hosts. The general goal is to protect the results within the chain of partial results from being modified. Given a sequence of partial results $m_0, ..., m_{n-1}$, then *forward integrity* is defined as follows:

*Definition 2 (weak forward integrity):* If $i_n$ is the first malicious agent place on the itinerary, the integrity of each partial result $m_0, ..., m_{n-1}$ is provided.

*Weak forward integrity* is conceptually not resistant to cooperating malicious hosts and agent places that are visited twice. To really protect the integrity of partial result we need a definition without constraints.

*Definition 2 (strong forward integrity):* None of the encapsulated messages $m_k$, with $k < n$, can be modified.

In this paper we will refer to *forward integrity* as to *strong forward integrity* (when applicable). To make notion of *forward integrity* more useful, we will define also *publicly verifiable forward integrity*, which enables any host to detect compromised agents:

*Definition 2 (publicly verifiable forward integrity):* Any host $i_n$ can verify that the chain of partial results $m(i_0), ..., m(i_{n-1})$ has not been compromised.

The other important notion concerning agent's integrity is a concept of *black-box security* ([12], [7]). Its main idea is to generate executable code from a given agent's specification that cannot be attacked by read (disclosure) or modification attacks. An agent is considered to be *black-box* if at any time the agent code cannot be attacked in the above sense, and if only its input and output can be observed by the attacker.

## 3. Related work

There are two main concepts for protecting mobile agent's integrity:
- detection or prevention of tampering,
- providing trusted environment for agent's execution.

The second group of methods is more concentrated on the whole agent system than on an agent in particular. These seem to be easier to design and implement but, as presented in [12], mostly leads to some problems. The idea that agent works only with a group of trusted hosts makes the agent less mobile than it was previously assumed. Also an agent may need different levels of trust (some information should be revealed to host but in another situation should be kept secret). Sometimes, it is not always clear in advance that current host is trusted.

Another way to provide such an environment is special tamper-resistant hardware, but the cost of such a solution is still very high.

In this paper will concentrate on the "built-in" solutions because they enable agent to stay mobile in a strong sense (as presented in the Section 2) and still provide the agent with mechanisms to detect or prevent tampering.

Detection implies that the technique is aimed at discovering unauthorized modification of the code or the state information. Prevention implies that the technique is aimed at keeping the code and the state information from being changed in any way. To be effective, detection techniques are more likely than prevention techniques to depend on a legal or other social framework. The distinction between detection and prevention can be arbitrary sometimes, since prevention often involves detection ([9]).

## 3.1 Encrypted Functions

The *Encrypted Functions* (*EF*) is one step forward in implementing *the perfect black-box security*. It has been proposed initially in [14]. Since then other similar solutions were introduced ([1], [2], [3], [15]) and the method is believed to be a one of canonical solutions for preserving agent's integrity ([9], [12]).

The goal of *Encrypted Functions* [9] is to determine a method which will enable the mobile code to safely compute cryptographic primitives, such as the digital signature, even though the code is executed in non-trusted computing environments and operates autonomously without interactions with the home platform. The approach is to enable the agent platform to execute a program assimilating an encrypted function without being able to extract the original form. This approach requires differentiation between a function and a program that implements the function.

The *EF* system is described as follows ([12]):

$A$ has an algorithm to compute function $f$. $B$ has an input $x$ and is willing to compute $f(x)$ for $A$, but $A$ wants $B$ to learn nothing substantial about $f$. Moreover, $B$ should not need to interact with $A$ during the computation of $f(x)$.

To implement the system defined above, we must assume that the function $f$ can be encrypted into some other function $E(f)$. Then, the scheme can be constructed as follows:
- $A$ encrypts $f$ and obtains $E(f)$,
- $A$ creates program $P(E(f))$ (that implements $E(f)$,
- $A$ sends $P(E(f))$ to $B$,
- $B$ executes $P(E(f))$ on $x$,
- $B$ sends the results of program $(P(E(f))(x))$ to $A$,
- $A$ decrypts the received results and obtains $f(x)$.

The function $f$ can be, e.g., a signature algorithm with an embedded key or an encryption algorithm containing the one. This would enable the agent to sign or encrypt data at the host without revealing its secret key.

Although the idea is straightforward, it is hard to find the appropriate encryption schemes that can transform arbitrary functions as showed. The techniques to encrypt rationale functions and polynomials were proposed. Also the solution based on a RSA cryptosystem was described ([3]).

## 3.2 Time Limited Black-box Security and Obfuscated Code

In the previous section we introduced a notion of *a black-box security*. Since it is not possible to implement it today, the relaxation of this notion was introduced ([12]): it is not assumed that the black-box protection holds forever, but only for a certain known time. According to this definition, an agent has *the time-limited black-box property* if for a certain known time it cannot be attacked in the above-mentioned sense.

The central idea of this approach is to generate an executable agent from a given agent specification which cannot be attacked by read or manipulation attacks ([6]). The *time limited black-box* fulfills two *black-box* properties for this limited time:
- code and data of the agent specification cannot be read
- code and data of the agent specification cannot be modified

This scheme will not protect any data that is added later, although the variables that exist will be changeable.

In order to achieve *the black-box property*, several conversion algorithms were proposed. They are also called obfuscating or mess-up algorithms. These algorithms generate a new agent out of an original agent which differs in code but produce the same results.

The *code obfuscation* methods make it more complicated to obtain the meaning from the code. To change a program code into a less easy "readable" form they have to work in an automatic and parametric manner. The additional parameters should make possible that the same original program is transformed into different obfuscated programs. The difficulty is to transform a program in a way that the original (or a similar easily understandable) program cannot be re-engineered automatically. Another problem is that it is quite difficult to measure the quality of obfuscation, as this not only depends on the used algorithm but on the ability of the re-engineer as well.

Since an agent can become invalid before completing its computation, obfuscated code is suitable for applications that do not convey information intended for long-lived concealment. Also it is still possible for an attacker to read and manipulate data and code but as the role of these elements cannot be determined, the results of this attack are random and have no meaning for the attacker.

### 3.3 Cryptographic Traces

Giovanni Vigna introduced *cryptographic traces* (also called execution traces) to provide a way to verify the correctness of the execution of an agent ([17], [16]). The method is based on traces of the execution of an agent, which can be requested by the originator after the agent's termination and used as a basis for execution verification. The technique requires each platform involved to create and retain a non-repudiation log or trace of the operations performed by the agent while resident there, and to submit a cryptographic hash of the trace upon conclusion as a trace summary or fingerprint. A trace is composed of a sequence of statement identifiers and platform signature information. The signature of the platform is needed only for those instructions that depend on interactions with the computational environment maintained by the platform. For instructions that rely only on the values of internal variables, a signature is not required and, therefore, is omitted.

This mechanism allows detecting attacks against code; state and control flow of mobile agents. This way, in a case of tampering, the agent's owner can prove that the claimed operations could never been performed by the agent.

The technique also defines a secure protocol to convey agents and associated security related information among the various parties involved, which may include a trusted third party to retain the sequence of trace summaries for the agent's entire itinerary. If any suspicious results occur, the appropriate traces and trace summaries can be obtained and verified, and a malicious host identified.

The approach has a number of drawbacks, the most obvious being the size and number of logs to be retained, and the fact that the detection process is triggered sporadically, based on suspicious results' observations or other factors. Other more subtle problems identified include the lack of accommodating multi-threaded agents and dynamic optimization techniques. While the goal of the technique is to protect an agent, the technique does afford some protection for the agent platform, providing that the platform can also obtain the relevant trace summaries and traces from the various parties involved.

### 3.4 Chained MAC protocol

Different versions of *chained MAC protocol* exist ([6]). Some of them require existence of public key infrastructure, other are based on a single key. This protocol enables an agent to achieve full forward integrity. To utilize this protocol only the public key of the originator has to be known by all agent places. This can be imagined in a scenario where the originator is a rather big company that is known by its smaller suppliers.

Assume that $r_n$ is a random number that is generated by each host. This value will be used as a secret key in a *Message Authentication Code*. The partial result $m_n$ (single piece of data, generated on $n^{th}$ host, and see Section 2), the random seed $r_n$ and the identity of the next host are encrypted with the public key of the originator $K(i_0)$, forming the encapsulated message $M_n$:

$$M_n = \{r_n, m_n id(i_{n+1})\}_{K_{i_0}}$$

*A chaining relation* is defined as follows (*H* denotes here a hash-function):

$$h_0 = \{r_n, m_0 id(i_0)\}_{K_{i_0}}$$
$$h_n = H(h_{n-1}, r_n, o_n, id(i_{n+1})).$$

When an agent is migrating from host $i_n$ to $i_{n+1}$:

$$i_n \rightarrow i_{n+1} : \{M_k \mid 0 \leq k \leq n\}, h_n.$$

Similar schemes are also called *Partial Results Encapsulation* methods ([9]).

### 3.5 Watermarking

Watermarking is mainly used to protect the copyrights for digital contents. A distributor or owner of the content embeds a mark into a digital object, so its ownership can be proved. This mark is usually secret. Most methods exploit information redundancy. Some of them can also be used to protect the mobile agent data and code.

One of methods of watermarking is proposed in [5]. A mark is embedded into the mobile agent by using software watermarking techniques. This mark is transferred to the agent's results during the execution. For the executing hosts the mark is a normal part of results, is "invisible". If the owner of agent detects that mark has been changed (it is different from expected) than he has a proof that the malicious host was manipulating the agent data or code.

The paper presents three ways of embedding the mark into the agent:
- marking the code,
- marking the input data,
- marking the obfuscated code.

The mark or marks are validated after the agent returns to its originator.

Possible attacks against this method include:
- eavesdropping: if the data is not protected in any way (e.g. not encrypted) it can be read by every host
- manipulation: the malicious host can try to manipulate either the agent's code or data to change the results and still keep the proper mark.
- collusion: a group of malicious hosts can cooperate to discover the mark by comparing the obtained results.

### *3.6 Fingerprinting*

Software fingerprinting uses software watermarking techniques in order to embed a different mark for each user. Software fingerprinting shares the same weaknesses than these of software watermarking: marks must resilient to manipulation and "invisible" for observers.

The method for fingerprinting was proposed in [5]. Contrary to the watermarking methods, presented previously, here the embedded mark is different for each host. When the agent returns to the owner, all results are validate. So the malicious host is directly traced.

In the mobile agent fingerprinting approach, the embedded mark is different for each host. The way that marks are embedded in the mobile agent watermarking approach can also be used in the mobile agent fingerprinting.

The difference between mobile agent watermarking and fingerprinting is the fact that it is possible to detect collusion attacks performed by a group of dishonest hosts.

The paper presents three ways of embedding the mark into the agent:
- marking the code: in this case, malicious hosts have the possibility of comparing their different codes in order to locate their marks.
- marking the input data: the data are usually different for each host, so it is harder to identify the mark.

The procedure is similar to the mobile agent watermarking approach. However, the origin host must know what each mark for each host and their location. One of possibilities of reconstructing the marks can be catching the information of the previously chosen places in the results.

Possible attacks against this method include:
- eavesdropping: if the data is not protected in any way (e.g. not encrypted) it can be read by every host
- manipulation: the malicious host can try to manipulate either the agent's code or data to change the results and still keep the proper mark.
- collusion: colluding hosts comparing their data or results cannot extract any information about the mark, because all hosts have a different input data and a different embedded mark.

## 3.7 Other protocols

*Publicly Verifiable Chained Digital Signatures*
    This protocol allows verification of the agent's chain of partial results not only by the originator, but also by every agent place. However, it is still vulnerable to interleaving attacks. This protocol makes it possible that every agent place that receives an agent can verify that it has not been compromised. This saves computing power in the case an agent has indeed been compromised because the agent place reasonably can refuse to execute a compromised agent.

*Environmental Key Generation*
    This scheme allows an agent to take predefined action when some environmental condition is true. The approach centers on constructing agents in such a way that upon encountering an environmental condition (e.g., via a matched search string), a key is generated, which is used to unlock some executable code cryptographically. The environmental condition is hidden through either a one-way hash or public key encryption of the environmental trigger. The technique ensures that a platform or an observer of the agent cannot uncover the triggering message or response action by directly reading the agent's code.

*Itinerary Recording with Replication and Voting*
    A faulty agent platform can behave similarly to a malicious one. Therefore, applying fault tolerant capabilities to this environment should help counter the effects of malicious platforms. One technique of such a kind for ensuring that a mobile agent arrives safely at its destination is through the use of replication and voting. The idea is that rather than using a single copy of an agent to perform a computation, multiple copies are used. Although a malicious platform may corrupt a few copies of the agent, enough replicas avoid the encounter to successfully complete the computation.

## 4. Cryptographic primitives

We utilized two cryptographic primitives in the proposed scheme:
- a zero-knowledge proof (in a form of an identification protocol)
- a secure secret sharing scheme.

Below is a short description of the protocols utilized.

## 4.1 Zero-knowledge proofs

Zero knowledge proof system ([13]) is a protocol which enables one party to *prove* the possession or knowledge of a "secret" to the other party, without revealing anything about it, in the information theoretical sense. These protocols are also known as minimum disclosure proofs. Zero knowledge proofs involve two parties: the prover who possesses a secret and wishes to convince the verifier (the second party), that he indeed has a secret. The protocol is realized as an interaction between the parties. At the end of the protocol, the verifier should be convinced only if the prover knows the secret. If, however, the prover does not know it, the verifier will be sure of it with an overwhelming probability.

The zero-knowledge proof systems are ideal for constructing identification schemes. A direct use of a zero-knowledge proof system allows unilateral authentication of P (Peggy) by V (Victor) and require a large number of iterations, so that verifier knows with an initially assumed probability that prover knows the secret (or has the claimed identity). This can be translated into the requirement that the probability of false acceptance be $2^{-t}$ where $t$ is the number of iterations. A zero knowledge identification protocol reveals no information about the secret held by the prover, under some reasonable computational assumptions.

## 4.2 Secure secret sharing scheme

A $(t, n)$ threshold secret sharing scheme ([13]) distributes a secret among $n$ participants in such a way that any $t$ of them can recreate the secret. But any $t-1$ or fewer members gain no information about it. The piece held by a single participant is called a *share* or *shadow* of the secret. Secret sharing schemes are set up by a trusted authority - called a dealer who computes all shares and distributes them to participants via secure channels. The participants hold their shares until some of them decide to combine their shares and recreate the secret. The recovery of the secret is done by the so-called combiner who on behalf of the co-operating group computes the secret. The combiner is successful only if the reconstruction group has at least $t$ members.

*Definition 5:* Assume that secret belongs to the set $K$ and shares are from the set $S$. A $(t, n)$ threshold scheme is a collection of two algorithms.

The first algorithm called the dealer,
$$D : K \rightarrow S_1 \times S_2 \times ... \times S_n$$
assigns shares to the participants for a random secret $k \in K$. The participant $P_i \in P$ gets his/her share $s_i \in S_i$. If all share sets $S_i$ are equal we simply say that $s_i \in S$. The second algorithm (the combiner)
$$C : S_{i_1} \times S_{i_2} \times ... \times S_{i_j} \rightarrow K$$
takes shares and computes the secret. The combiner recovers the secret only if the number $j$ of different shares is equal to or bigger than $t$ ($j \geq t$). It fails if the number $j$ of shares is smaller than $t$ ($j < t$).

## 5. New concept of the integrity protection

In the proposed system we assume that there exist at least three parties:
- a manager,
- an agent,
- a host.

The manager can be an originator of an agent. It plays a role of a verification instance in the scheme and creates initial countermeasures for the agent. The manager also plays a role of a *Trusted Third Party*.

### *5.1 Basic idea*

The zero-knowledge proof systems enable verifier to check validity of the assumption that the prover knows a secret. In our system the verifiers would be the manager or owner of agents and, obviously, agents would be the provers. In the initial phase, manager computes set of secrets. The secrets are then composed into the agent, so that if manager asks an agent to make some computations (denote them as a function *f*), the result of this would be a valid secret. This function should have the following property:
- if we have $x_1$ and $f(x_1)$ than it is computationally infeasible to find such $x_2$ that $f(x_2) = f(x_1)$.

If the secret is kept within an agent, than also the host can use zero-knowledge protocol to verify it. Every authorized change of agent's state results in such a change of the secret that it remains valid. On the other hand, every unauthorized change leads to loosing the secret - so in the moment of verification by host or manager, the agent is not able to prove possession of a valid secret. In our system the host can tamper an agent and try to make such changes that he will be still able to obtain a proper secret, but the characteristics of function *f* will not allow doing this. A possible candidate for the function *f* can be a hash function. Our approach is a detection rather than prevention.

## 6. The protocol

Our protocol is not directly based on the complete zero-knowledge proof, but on the particular identification system based on zero-knowledge proof. We choose the Guillou-Quisquater (GQ) scheme ([8]) as the most convenient for our purposes. In this scheme the manager has a pair of RSA-like keys: a public $K_P$ and a private one $k_p$. The manager also computes the public modulus $N = p \cdot q$, where *p, q* are RSA-like primes. The following equation has to be true:
$$K_P \cdot k_p \equiv 1 \bmod \varphi(N),$$
where $\varphi(N)$ is the value of Euler function of *N*. The pair $(K_P, N)$ is made public.

### *6.1 The initial phase*

The initial phase has three steps:
1. The manager computes set of so-called identities, denoted as $ID_p$ and their equivalences denoted as $J_p$. It does not matter how $J_p$ is obtained if it is obvious for all participants how to obtain $J_p$ from $ID_p$. The pairs ($ID_p$, $J_p$) are public and can be distributed among hosts. The manager computes a secret value for each $ID_p$:
$$\sigma_p = J_p^{-k_p} \bmod N$$

The $\sigma_p$ is a secret that will be "hidden" in an agent.

2. The $\sigma_p$ should be "composed" or "built" into an agent. To do this we utilize the Asmuth and Bloom secure secret sharing scheme ([13]). The manager randomly chooses $m$ prime or co-prime numbers (called public modulus):

$$p_j, j = 1,2,...,m, p_0 < ... < p_j < ... < p_m.$$

They are publicly known. Then the manager (playing a role of a dealer in the secret sharing scheme) instead of selecting at random an integer $s$, such as

$$s < \prod_{j=1}^{t} p_j$$

he computes it, preserving following conditions:

$$s \bmod p_0 \equiv \sigma_p$$

and

$$p_m < s < \prod_{j=1}^{t} p_j$$

After computing $s$ the manager creates also appropriate shares:

$$s_i = s \bmod p_i.$$

Then, the $t-1$ shares are composed into agent and the rest is distributed among the hosts via a secure channel.

3. The manager now needs to compose the shares into an agent in a way that when the agent is in a proper execution state, he is able to obtain from his code/state variables the correct shares. Since the agent is still a computer program he can be described as a *Finite State Machine*. So the shares can be connected to a certain state in which the agent currently is: the proper execution states will generate correct shares, while others not. To create the shares, the hash function, maybe based on some internal variables can be used. Alternatively, an encryption function with a manager's public key can be used.

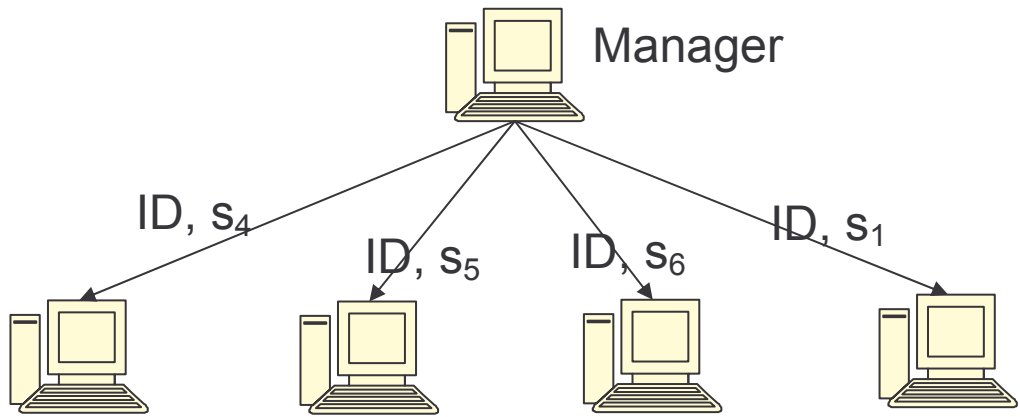

**Fig. 1 Distributing ID and shares to hosts**

## 6.2 The first scenario: a host validating an agent

The figure below shows general steps of situation that a host checks that just migrated agent is valid.

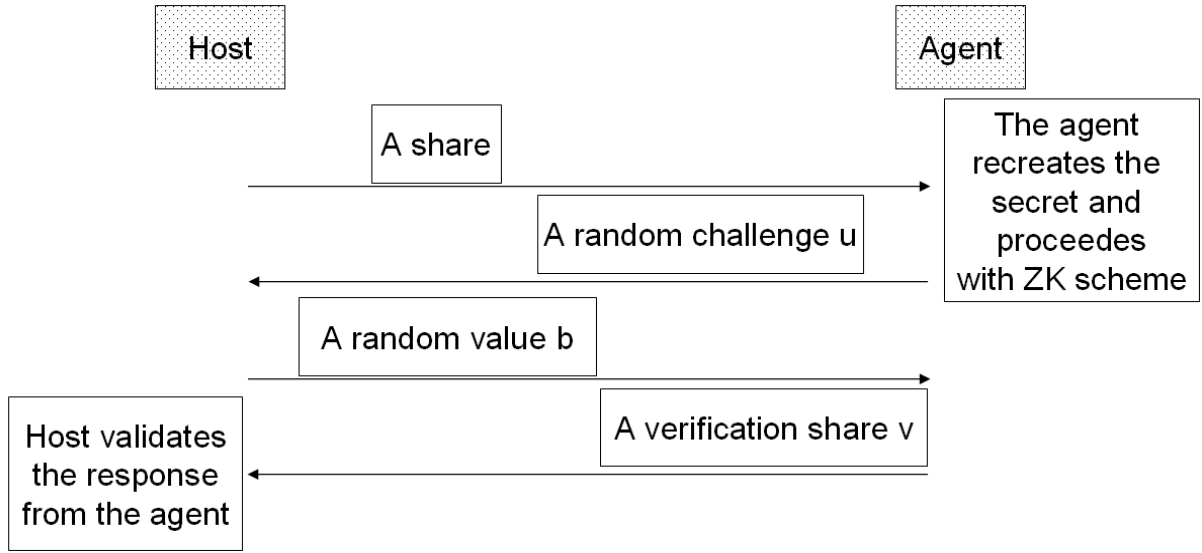

**Fig. 2 The verification of an agent's integrity**

1. The host wanting to verify an agent's integrity sends him his share $s_h$.
2. The agent creates the rest of the shares from his code and the execution state. He recreates the secret (playing a role of a combiner for the Asmuth and Bloom secure secret sharing scheme) by solving the following system of equations:

$$s_{i_1} \equiv s \bmod p_{i_1}$$
$$\ldots\ldots$$
$$s_{i_t} \equiv s \bmod p_{i_t}$$

   This system has a unique solution according to the Chinese Reminder Theorem. The agent computes the secret $\sigma$ and uses it for the rest of the scheme, which is a zero-knowledge proof based identification protocol.
3. The agent sends the host a challenge: a number computed based on a random value $r$, $r \in \{1,...,N-1\}$. It is computed as following:

$$u \equiv r^{K_P} \bmod N$$

4. After receiving the challenge the host chooses a random value $b \in \{1,...,N\}$ and sends it to the agent.
5. The agent computes next value ($v$) basing on the number from the host and on agent's secret value $\sigma$:

$$v \equiv r \cdot \sigma^b \bmod N$$

6. The host uses information received from the manager, $ID_p$ to obtain $J_p$ and verifies if $v$ is a proper value. To validate the response from the agent, the host checks if

$$J_P^b \cdot v^{K_P} \equiv u \bmod N$$

   If the equation is true than the agent proved that he knows the proper secret and neither his code nor execution state were changed.

The manager can compute many identities, which may be used with different execution states. In that situation the agent should first inform host which identity should be used, or host can try to validate the received value $v$ for all possible identities. This 2nd part of this protocol, starting from the agent sending a challenge to the host, can be repeated to minimize the probability of not detecting any manipulation in the agent's code.

## 6.3 The second scenario: securing the data obtained by an agent

A similar scenario can be used to provide integrity to the data obtained by the agent from different hosts. A malicious host could try to manipulate the data delivered to agent by the previous hosts. To ensure that this is not possible, the agent can use the zero-knowledge protocol to protect the data. For each stored data *d*, agent can choose at random $r \in \{1,...,N-1\}$ and compute

$$v \equiv r \cdot \sigma^d \mod N$$

Then manager can verify by computing and comparing:

$$J_P^d \cdot v^{K_P} \equiv u \mod N .$$

That way for every received data *d* the agent would have a unique "proof" that the data was not manipulated.

## 7. Security of the proposed scheme

The proposed scheme should be used with more that one identity ($ID_p$). This would make possible to manipulate the code and the data very hard. The best approach is to use one secret for each host.

We assume that the malicious host is able to read and manipulate an agent's data and code. He can try to obtain from an agent's execution state the proper shares. He can also try to obtain a proper secret and manipulate the agent's state and variables in a way that the obtained secret would stay the same. But he does not know other secrets that are composed into the agents, also he does not know more shares to recreate those secrets, so, any manipulation would be detected by the next host.

Also even when host is able to recreate the current secret, he is not able to manipulate the data that was obtained by the agent earlier from other hosts. Since he cannot produce a valid secret $\sigma$ for given data *d*, he is not able to forge the *v*, the way that using a zero-knowledge proof would not reveal the changes. The proposed solution fulfills the *forward integrity* definition: none of data and corresponding *v* values can be changed without a future detection by the manager.

The protocol is not able to prevent any attacks that are aimed at destroying the agent's data or code, meaning that a malicious host can "invalidate" any agent's data. But this is always a risk, since the host can simply delete an agent.

## 8. Future work

One area for development is to find the most appropriate function for composing secrets into hosts: the proposed solution seems to fulfill the requirements, but some evaluation should be done.

One of the possibilities for a future work would be to integrate the proposed solution to some agent security architecture, possibly one that would also provide an agent with strong authentication methods and anonymity ([18]). Then, such a complex system could be evaluated and implemented.

## 9. Conclusions

This paper provides description of various protocols and methods for preserving the agent's integrity. The basic definitions and notions were introduced. The most important mechanisms are overviewed and discussed. We also propose a new concept for detection of a tempering an agent, based on a zero-knowledge proof system. The proposed scheme secures both an agent's execution state and internal data. The system requires some additional research and development work but it seems to be a promising solution to the problem of providing agent with effective countermeasures against attacks on the integrity.